\documentclass[12pt]{article}
\usepackage{graphicx}
\usepackage{amssymb}\usepackage{bm}\usepackage{amsfonts}
\newcommand{\beq}{\begin{equation}}\newcommand{\eeq}{\end{equation}}\newcommand{\beqa}{\begin{eqnarray}}
\newcommand{\eeqa}{\end{eqnarray}}\newcommand{\w}{\wedge}\newcommand{\ts}{\textstyle}
\newcommand{\nn}{\nonumber}

\begin{document}
%\maketitle
{\renewcommand{\thefootnote}{\fnsymbol{footnote}}
\hfill  IGC--08/5--1\\
\medskip
\begin{center}
{\LARGE  A Mesoscopic Quantum Gravity Effect}\\
\vspace{1.5em}
Andrew Randono\footnote{e-mail address: {\tt arandono@gravity.psu.edu}}
\\
\vspace{0.5em}
Institute for Gravitation and the Cosmos,\\
The Pennsylvania State
University,\\
104 Davey Lab, University Park, PA 16802, USA\\
\vspace{1.5em}
\end{center}
}
\abstract{We explore the symmetry reduced form of a non-perturbative solution to the constraints of quantum gravity corresponding to quantum de Sitter
space. The system has a remarkably precise analogy with the non-relativistic formulation of a particle falling in a
constant gravitational field that we exploit in our anaylsis. We
find that the solution reduces to de Sitter space in the semi-classical limit, but the uniquely quantum features of
the solution have peculiar property. Namely, the unambiguous quantum structures are neither of Planck scale
nor of cosmological scale. Instead, we find a periodicity in the volume of the universe whose period, using the observed
value of the cosmological constant, is on the order of the volume of the proton.}
\pagebreak
\section{Three roads to quantum gravity phenomenology}
In contrast to popular lore, quantum effects are manifest at all length scales. The quantum regime is not restricted
to the microscopic, but it is generically associated with extremes, e.g. high energy, low temperature, high density,
etc... This association is in part historical: had a given effect been discovered prior to the development of
quantum mechanics, it would likely be called classical. This is evidenced by the occasional unambiguous quantum
process that is categorized as a classical phenomenon, such as the theory of electrical conductivity whose
underlying mechanism is purely quantum mechanical yet the associated Ohm's law is as classical as they come. It
would be a happy occurrence if such were the case in quantum gravity---that a familiar, seemingly classical phenomenon
necessarily had its roots in quantum gravity. In this paper we address a slightly more modest proposal with this goal in
mind. In particular we draw attention to the emergence of intermediate length scale structures from canonical quantum
gravity via a particular solution to the non-perturbative quantum constraints.

To categorize our proposal, it will be useful to distinguish three broad classes of quantum gravity phenomenology: the
microscopic, the macroscopic, and the mesoscopic. Most discussions of quantum gravity phenomena focus on the Planck
scale set by
$\ell_{Pl}=\sqrt{\frac{G\hbar}{c^{3}}}\approx 10^{-35}m$. Due to the incredibly
small size of the Planck scale, direct observation of such quantum gravity effects is not likely any time soon so
realistic phenomenology must appeal to other length scales. Somewhat paradoxically, recurrent themes suggest
that quantum gravity effects might be manifest at macroscopic, cosmological scales as in the dualities, brane world
scenarios, and possible large extra dimensions of String Theory, or the quantum gravity inspired explanations for the smallness of the cosmological
constant\cite{Stephon:CCrelaxation}. This paper, however, concerns the possibility of an intermediate mesoscopic scale
emerging from quantum gravity.
Although we will stop short of a full analysis of possible mesoscopic physics with quantum gravitational roots, we
will see very clearly structures of mesoscopic scale emerging out of quantum de Sitter space. We will exploit a
remarkably precise analogy between a non-relativistic particle in free-fall and the Kodama state, which is a candidate solution to the constraints of non-preturbative quantum gravity corresponding to de Sitter space. The latter will inherit many of the interesting quantum features of the former.

\section{Particle in free-fall}
Let us start with a brief review of the quantum mechanical description of a particle in free-fall in a constant
gravitational
field underscoring the aspects of the solution that will carry over to quantum de Sitter space (for an excellent
pedagogical discussion including experimental consequences, see \cite{Backreaction:BouncingNeutrons}). 
In light of the weakness of the gravitational force (even at the surface of the earth when acting on subatomic
particles) it may come as a surprise to many that a particle in free-fall could exhibit observable quantum effects at
all. In reality, the unique quantum features of the distribution are considerably larger than one might generically expect.

\begin{figure}[h] %  figure placement: here, top, bottom, or page
   \centering
   \includegraphics[width=.6\textwidth]{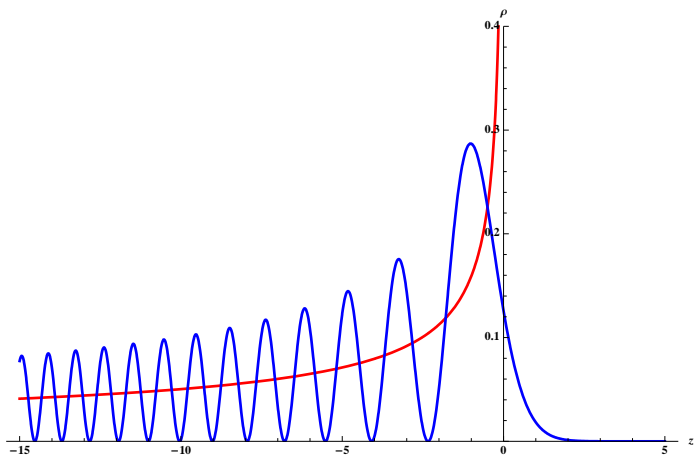} 
   \caption{\small{The classical probability density is shown in red and the quantum density shown in blue. We clearly see a close agreement between the two that is a consequence of the WKB nature of the state. The ``fringes" of the quantum state are evident in the oscilatory behavior of the wavefuntion.}}
   \label{freefall}
\end{figure}

The Schrodinger equation for a particle of mass $m$ in free-fall is
\beq
-\frac{\hbar^{2}}{2m}\frac{d^{2}\psi}{dz^{2}}+mg z\,\psi=E\,\psi\,.
\eeq 
This is a familiar differential equation whose unique bounded solution is 
\beq
\psi(z)=\mathcal{N}Ai\left(\left(\frac{2}{\lambda^{2}_{c}\lambda_{g}}\right)^{1/3}\left(z-z_{0}\right)\right)
\eeq
where $\mathcal{N}$ is a normalization constant, $z_{0}=E/mg$ is the peak of the classical trajectory,
$\lambda_{c}=\frac{\hbar}{mc}$ is the Compton wavelength, and $\lambda_{g}=\frac{c^{2}}{g}$ is a length scale set by the
macroscopic gravitational field. In the momentum representation, the wavefunction is pure phase, and it is both an
exact solution and a zeroeth order WKB state. In this sense, the quantum state is as classical as they come, and one
should expect a close agreement between the classical and quantum probability distributions. The classical probability
density at a point in phase space is proportional to
the amount of time spent in a small neighborhood of the point, which we can write, $\rho_{class}(z)\sim\sum_{crossings}
\frac{1}{|\dot{z}(z)|}=\frac{1}{\sqrt{\frac{g}{2}(z_{0}-z)}}$, summing over the number of times the particle
enters the small region. The asymptotic expansion of the Airy function reveals the characteristic exponential decay of
the quantum probability distribution outside the classically forbidden region, and the $1/\sqrt{z_{0}-z}$ behavior
within the classically allowed region away from the turning point. The classical and quantum probability distributions are shown superimposed in figure \ref{freefall} on page \pageref{freefall} where the classical--quantum correspondence is immediately evident. 

The uniquely quantum structures of the wavefunction come from a peculiar feature of the quantum wavefunction: whereas the
classical trajectory and the classical structure of the quantum wavefunction is independent of the mass (a consequence of the classical equivalence principle), the
oscillatory behavior of the wave function does depend on the mass. These quantum structures are scaled by length
parameter $(\lambda^{2}_{c}\lambda_{g})^{1/3}$. Although $\lambda_{c}$ is typically very small for subatomic particles,
$\lambda_{g}$ is very large and the combination results in quantum structures that are large enough to be observable.
The phase gives rise to ``fringes" resulting from
the oscillatory behavior of the quantum probability distribution and the exponentially damped tail in the classically
forbidden region. For a neutron near the surface of the Earth, this yields a width
for the first fringe of approximately $(\lambda^{2}_{c}\lambda_{g})^{1/3}\approx 10^{-5}m$, which is close to the
resolving power of the naked eye! For
smaller mass particles the fringe width can be much larger. For example, taking the mass limit on the electron neutrino
$m_{\nu_{e}}\lessapprox 2 eV$, the largest fringe size is on the order of one meter or more. The peculiar balance
between a very small and a very large quantity that gives rise to the intermediate quantum scale will carry
over to our cosmological model.

\section{Quantum de Sitter space and the Kodama state}
The Kodama state and its various generalizations have been argued to be non-perturbative solutions to the quantum
operator
analogues of the equations defining de Sitter space\cite{Smolin:kodamareview, Randono:GKI, Randono:GKII, Randono:Thesis}.
Just as with the particle in free-fall, the state can be viewed
as an exact state and a zeroeth order WKB state. The simplest route to the construction is to begin with the Einstein-Cartan action with a
cosmological constant, and for generality we will add a non-minimal parity violating term often referred to as the
Holst modification or Immirzi term\footnote{In the first line we have used the index free Clifford notation, which is
generally easier to work with for simple calculation. In this notation, the spin connection is valued in the Clifford
bi-vector algebra, $\omega=\frac{1}{4}\gamma_{[I}\gamma_{J]}\omega^{IJ}$, the tetrad is valued in the vector elements
of the Clifford algebra, $e=\frac{1}{2}\gamma_{I}e^{I}$, and the dual is
$\star=-i\gamma_{5}=\gamma^{0}\gamma^{1}\gamma^{2}\gamma^{3}$. Explicit wedge products have been dropped and the trace
over the Clifford matrices is assumed in the action.}:
\beq
S=\frac{1}{k}\int_{M}\star \,e \,e \left(R-\ts{\frac{\lambda}{6}}\,e\, e\right)-\ts{\frac{1}{\beta}}\,e \,e\, R
\eeq
de Sitter space is defined by the condition $R=\frac{\lambda}{3}\,e\,e$ (the Immirzi parameter has no effect at the
classical level). Inserting this
into the action and setting the three-torsion to zero (a second class constraint that emerges in the detailed constraint
algebra) we arrive at the Kodama state in the connection representation
\beq
\Psi_{R_{\Gamma}}[A]=\mathcal{N}e^{i S_{0}}=\mathcal{P} \exp\left[\frac{3i}{4k\lambda\beta^{3}}\int_{\Sigma}
Y[A]+2(1+\beta^{2})A\w R_{\Gamma}\right] \label{GKState1}
\eeq
where the implied trace is now in the adjoint representation of $SU(2)$ we have absorbed all terms that depend only on
the classical configuration $E$ into an overall phase factor. The author has argued that the above 
defines an auxiliary Hilbert space of states labelled by a particular configuration of the three-curvature
$R_{\Gamma}[E]$, and the unique diffeomorphism and gauge invariant state corresponding to $R_{\Gamma}=0$ is the
quantum version of de Sitter space in the flat $\mathbb{R}^{3}$ slicing.

We are primarily interested in the symmetry reduced version of the state. For simplicity we consider the Kodama state in
the limit that $\beta\rightarrow \infty$. As we will see the relevant quantum structures we will find are
well above the Planck scale, and the Immirzi parameter typically affects physics at the Planck scale, thus we are justified in
taking this limit.
Furthermore, preliminary investigations to be reported in a follow up paper suggest these stuctures are unnaffected by the introduction of the Immirzi parameter. 

Beginning with the Freidman-Robertson-Walker ansatz for the metric\footnote{In our conventions, 
the coordinates carry dimensions of length so the scale factor $a(\tau)$ is unitless.},
\beq
ds^{2}=-N^{2}d\tau^{2}+a^{2}\left(\frac{dr^{2}}{1-\kappa r^{2}}+r^{2}d\Omega^{2}\right)\,,
\eeq
the action reduces to:
\beqa
S&=& \frac{3 L^3}{8\pi G}\int_{\mathbb{R}}dt\left(a^{2}\,\ddot{a}+a\,\dot{a}^{2}+\kappa a
-\ts{\frac{\lambda}{3}}\,a^{3}\right)\nn\\
&=& \frac{3}{8\pi G}\int_{\mathbb{R}}dt\left(\mu \,\dot{k} +N\sqrt{\mu}(k^{2}+n-\ts{\frac{\lambda}{3}}\,\mu)\right)
\eeqa
where, ${L}^{3}$ is a fiducial
volume of the cell over which the action is evaluated,
and we have defined $\mu\equiv L^{2}a^{2}$,
$k\equiv L\dot{a}$, and $n\equiv L^{2}\kappa$. The last variable, $n$, is positive, zero, or negative
corresponding to a closed spherical, open flat, or open hyperbolic three geometry respectively.
From the form of the action, we
identify the fundamental Poisson bracket, $\{k,\mu\}=\frac{8\pi G}{3}$, which carries over to the operator commutator
\beq
[\,\hat{k}\,,\,\hat{\mu}\,]=i\frac{8\pi G}{3}. 
\eeq
Classically, there is only one solution to the Hamiltonian constraint (assuming the three-metric is non-degenerate),
and it is the defining condition of de Sitter space in symmetry reduced variables:
\beq
n+k^{2}=\ts{\frac{\lambda}{3}}\mu\,. \label{dSConditionReduced}
\eeq
As previously, we insert this solution back into the action to arrive at the WKB solution in the $k$-representation:
\beq
\Psi[k]=\mathcal{P} e^{iS_{0}[k]}=\mathcal{P}\exp\left[\frac{9i}{8\pi G\lambda}\left(\frac{1}{3}k^{3}+kn\right)\right]\,.
\label{GKReduced1}
\eeq
A similar form for the Kodama state was also obtained in the context of symmetry reduced Plebanski theory in \cite{Perez:BFKodama}. It can easily be verified that the above state is simply the symmetry reduced form of (\ref{GKState1}) in the limit that
$\beta\rightarrow \infty$. We note the integer $n$ in the symmetry reduced state plays the role of the curvature
parameter $R_{\Gamma}$.

The wavefunction is easier to intrerpret in the $\mu$-representation. The Fourier transform of
(\ref{GKReduced1}) is the bounded solution to the Airy differential equation yielding:
\beq
\Psi(\mu)=\mathcal{N}Ai\left(-\left(\frac{3}{8\pi \,\ell^{2}_{pl} r_{0}}\right)^{2/3}
\left(\mu-n r^{2}_{0}\right)\right)\label{GKReduced2}
\eeq
where $r_{0}=\sqrt{\frac{3}{\lambda}}$ is the de Sitter radius.
The semi-classical analysis of this state follows closely with that of a particle in free-fall. 

\section{Semi-classical analysis of the state}
To carry out the semi-classical analysis of the wavefunction we first need to identify a time variable. In fact, we
have already implicitly chosen one.
To see this, recall that the symmetry reduced Kodama state
is in the kernel of the quantum operator version of the de Sitter condition (\ref{dSConditionReduced}). However, the
Hamiltonian constraint is $C_{H}(N)=N\sqrt{\mu}(k^{2}+n-\ts{\frac{\lambda}{3}}\mu)$. If the Kodama state is to be in
the kernel of the Hamiltonian constraint, it is natural to choose a dynamical lapse, $N= \frac{\alpha}{\sqrt{\mu}}$,
where $\alpha$ is a constant so that the Hamiltonian operator is now precisely the operator corresponding to (\ref{dSConditionReduced})\footnote{As
always, subject to a particular (very natural) choice of operator ordering.}
Classically this corresponds to a choice of a non-standard time variable where the de Sitter solution now takes the form
\beqa
k(\tau)&=& \tau/r_{0} \nn\\
\mu(\tau)&=&\tau^{2}+ n r_{0}^{2}\,.
\eeqa
Thus, this choice for the lapse has effectively stretched the time variable such that the de Sitter trajectory is parabolic as opposed to hyperbolic. 

Now, given this trajectory, consider the classical probability distribution in the $\mu$-representation. The classical
probability density is $\rho_{class}(\mu)\sim  \sum_{crossings}\frac{1}{|\mu'(\mu)|}=
\frac{1}{|\sqrt{\mu-nr_{0}^{2}}|}$
which holds for $\mu \geq n r_{0}^{2}$ and $\rho_{class}=0$ for $\mu<n r_{0}^{2}$. The probability density blows up as
we approach $n r^{2}_{0}$ from above because the effective ``velocity" goes to zero at this point. This is the
analogue of the classical turning point of the particle in free-fall, corresponding to the throat of de Sitter space in
the $n=+1$
model where the universe reverses its contraction and begins to expand. Just as with the particle in free-fall, we have a very close match between the classical
and quantum probability density, $\rho_{quantum}=|\Psi(\mu)|^{2}$. The
asymptotic expansion of the Airy function yields\footnote{A similar asymptotic expansion for a candidate de Sitter solution involving a cosine rather than a sine was found using different methods long ago by Hawking \cite{Hawking:BigBangBook}.}:
\beqa
\rho_{quant}(\mu)& \stackrel{\mu\gg n r^{2}_{0}}{\sim}&
\frac{\sin^{2}\left(\frac{1}{4\pi G\hbar r_{0}}(\mu-n r_{0})^{3/2}+\frac{\pi}{4}\right)}
{\sqrt{(\mu-n r^{2}_{0})}}\nn\\
\rho_{quant}(-\mu) & \stackrel{\mu\gg -n r^{2}_{0}}{\sim}& 
\frac{\exp\left(-\frac{1}{2\pi G\hbar r_{0}} (\mu-n r_{0})^{3/2}\right)}{\sqrt{(\mu-n r^{2}_{0})}}\,.\label{KSAsymptotics}
\eeqa
Again we see the characteristic exponential decay of the wavefunction outside the classically forbidden region, and the purely quantum
oscillatory feature of the quantum probability density superimposed on the classical distribution inside the
classically allowed region.

\subsection{A mesoscopic length scale}
As with the particle in free-fall, the uniquely quantum structures depend on the balance of a very small and a very
large length scale.
Just as the
kinematics of the particle in free-fall is classically independent of the mass, the de Sitter solution is classically
independent of Newton's constant. However, the Kodama state does depend on $G=\ell_{pl}^{2}$, just as the the quantum
free-fall state
does depend on $m$. Consider the asymptotic expansion, (\ref{KSAsymptotics}), of the quantum probability
distribution in the classically allowed region. We clearly see an oscillatory structure superimposed on the classical
probability distribution. For large $\mu$, the distribution is oscillatory with respect to $\mu^{3/2}$. Recalling
that the physical volume (in the closed model) $2\pi^{2} \mu^{3/2}$, the quantum probability distribution has a
periodicity in the volume given by:
\beq
\Delta V=8\pi^{4}\ell_{pl}^{2}\,r_{0}.
\eeq
Thus, the scale of the uniquely quantum
features of the wavefunction are neither of Planck scale nor cosmological scale, but they reside in an intermediate,
mesoscopic scale. Using the observed value of the cosmological constant today, $\lambda\approx
10^{-120}/ \ell_{pl}^{2}\approx 10^{-50}m^{-2}$, the periodicity in the volume is approximately:
\beq
\Delta V\approx 10^{-42}m^{3}
\eeq
or converting this to a length scale:
\beq
(\Delta V)^{1/3}\approx 10^{-14}m \approx 10\times d_{proton}
\eeq
where $d_{proton}$ is the diameter of the proton. Thus, just as with the case of the free-fall quantum state, the
balance of microscopic and macroscopic length scales conspire to produce a mesoscopic quantum scale. Since the scale
of the pure quantum structures of the Kodama state is on the length scale of the strong interaction, the possibility
remains that a signature of these structures may be seen in, for example the relative abundance of matter in the early
universe.

To gain further insight into the nature of the quantum oscillations of the wavefunction, we construct an effective
spacetime by identifying the quantum probability distribution as an effective probability in the WKB analysis. That
is, we identify:
\beq
\rho_{effective}=|\psi|^{2} \stackrel{effective}{\approx}\frac{2}{|\mu'(\mu)|}\,.
\eeq
This is to be viewed as an effective equations for deducing $\mu$ as a function of $\tau$.
Rearranging terms we have $\int d\tau =\pm \int d\mu \,\rho_{effective}(\mu)
$ so that :
\beqa
\tau &=& \pm \mathcal{N}^{2}\int d\mu \left(Ai\left(-\alpha\left(\mu-n r^{2}_{0}\right)\right)\right)^{2}\nn\\
&=& \pm \mathcal{N}^{2}\left((\mu-n r^{2}_{0})Ai[-\alpha(\mu-n r^{2}_{0})]^{2}+\frac{Ai'[-\alpha(\mu-n
r^{2}_{0})]^{2}}{\alpha}\right)\equiv f(\mu)
\eeqa
where $\alpha=\left(\frac{3}{8\pi \ell^{2}_{pl}r_{0}}\right)^{2/3}$
The function $f(\mu)$ is invertible, giving an effective trajectory for $\mu$ as a function of time:
\beq
\mu_{effective}(\tau)=f^{-1}(|\tau |)\,.
\eeq
This trajectory is plotted below in figure \ref{QdeSitter1} on page \pageref{QdeSitter1} with the classical trajectory superimposed.
\begin{figure}[h]
\begin{center}
\includegraphics[width=.75\textwidth]{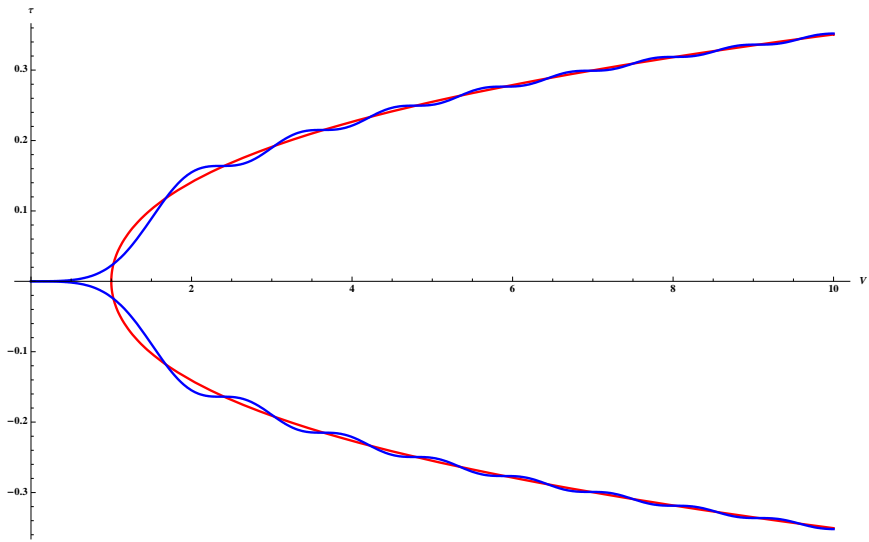}
\caption{\small{The red curve is the classical trajectory and the blue curve is the effective trajectory of the volume, $V$,
plotted as a function of $\tau$ on the vertical axis for the values $nr^{2}_{0}=1$ and $\left(\frac{3}{8\pi
\ell^{2}_{pl}r_{0}}\right)^{2/3}=3$. We clearly see a strong agreement between the classical and quantum trajectories.
The quantum trajectory appears to evolove via a series of quasi-discrete jumps with value $\Delta
V=8\pi^{4}\ell_{pl}^{2}\,r_{0}$.}}
\label{QdeSitter1}
\end{center}
\end{figure}

 The WKB analysis is not
valid near the classical turning point so the peculiar behavior of the effective scale factor at $\tau=0$ can be
discarded. The volume appears to evolve by a series of quasi-discrete jumps that cycle average to reproduce the
classical trajectory. It should be stressed that the quantum trajectory plotted in figure \ref{QdeSitter1} is only an effective trajectory---a proper treatment would require embedding the symmetry reduced de Sitter space in a full Hilbert space, identifying an internal time variable and plotting the expectation value of the volume in an appropriate state as a function of the internal time variable. This analysis is forthcoming in a follow-up paper.

\section{Concluding Remarks}
We have clearly seen a mesoscopic scale emerging from the non-perturbative description of quantum de Sitter space.
Furthermore, the essential features followed from a WKB analysis, and regardless of the details of the
quantization procedure one uses the WKB approximation should be valid in an appropriate regime. Thus, regardless of
the details of the quantum theory at the Planck scale, since the quantum structures of interest are much larger we
expect that they will remain. The numerical coincidence that these
fluctuations are on the order of the scale set by the strong interaction opens up the exciting possibility that this
type of quantum gravity effect might have consequences for the dynamics of the matter content of the universe. It remains to be seen whether this scaling has observational consequences. 

\section*{Acknowledgments}
I would like to thank Abhay Ashtekar and especially Golam Hossain for many stimulating discussions concerning this work and our collaborative follow-up paper.
\bibliography{MesoQGGRFV1}

\end{document}